\documentclass[aps,prb,preprint,showpacs]{revtex4}
\usepackage{bm}
\usepackage{graphicx}

\begin{document}
\title{Spin relaxation of two-dimensional holes in
strained asymmetric SiGe quantum wells}
\author{B.~A.~Glavin}
\affiliation{V.~E.~Lashkar'ov Institute of Semiconductor Physics,
Pr.~Nauki 41, Kiev 03028, Ukraine}
\author{K.~W.~Kim}
\affiliation{Department of Electrical and Computer Engineering,
North Carolina State University, Raleigh, NC 27695-7911, USA}
\begin{abstract}
We analyze spin splitting of the two-dimensional hole spectrum in
strained asymmetric  SiGe quantum wells (QWs). Based on the
Luttinger Hamiltonian, we obtain expressions for the
spin-splitting parameters up to the third order in the in-plane
hole wavevector. The biaxial strain of SiGe QWs is found to be a
key parameter that controls spin splitting.  Application to SiGe
field-effect transistor structures indicates that typical spin
splitting at room temperature varies from a few tenth of meV in
the case of Si QW channels to several meV for the Ge counterparts,
and can be modified efficiently by gate-controlled variation of
the perpendicular confining electric field. The analysis also
shows that for sufficiently asymmetric QWs, spin relaxation is due
mainly to the spin-splitting related D'yakonov-Perel' mechanism.
In strained Si QWs, our estimation shows that the hole spin
relaxation time can be on the order of a hundred picoseconds at
room temperature, suggesting that such structures are suitable for
p-type spin transistor applications as well.
\end{abstract}


\maketitle

\section{Introduction}

Recently, there has been considerable interest in the field of
 spintronic applications. They range from various
approaches in logic and memory elements to quantum
computation.\cite{spinel} Many spintronic device concepts rely on
the so-called Das-Datta spin transistor, \cite{dasdatta} where the
transfer of spins between the spin-polarized source and drain
contacts is controlled by the gate bias through the perpendicular
(i.e., confining) electric field in the quantum well (QW) channel.
Experimentally, the efforts have concentrated mainly on $n$-type
III-V devices; in particular, those with InGaAs QWs. The reasons
for this preference are (1) easy control of electron spin states
in this material\cite{ingaas-gate} and (2) availability of III-V
diluted magnetic semiconductors.\cite{ingaas-dms} Much less
attention has been paid to the hole-based spintronic applications, whose
rationale actually originates from the considerations related to
the bulk semiconductors. Indeed, in bulk cubic semiconductors the
top of the valence band at the Brillouin zone center is four-fold
degenerate (including spin), and hybridization of the hole states
for finite wavevectors depends on the direction of the wavevector.
As a result, hole scattering is supplemented by simultaneous
change of spin, and the so-called Elliott-Yafet mechanism of spin
relaxation is very effective.\cite{yafet} In fact, the hole spin
relaxation rate is close to the momentum relaxation rate; i.e.,
nonquilibrium spin relaxes too fast to allow control in any
realistic device. This picture is not exactly correct in the case
of two-dimensional (2D) holes confined in a QW, where the
degeneracy at the zone center is partially lifted due to
quantization. Moreover, the hole spectrum in an asymmetric QW is
completely nondegenerate for nonzero wavevectors,
qualitatively resembling that of electrons. The effective Hamiltonian for 2D
holes in GaAs QWs were analyzed, for example, in
Ref.~\onlinecite{wink}.

In this paper, we address the problem of the hole spectrum and spin
relaxation in strained SiGe QWs.  Previously, spin splitting in
such QWs was estimated based on the observation of the circular
photogalvanic effect\cite{holephotogal} and transport
measurements.\cite{transport} In Ref.~\onlinecite{holephotogal},
possible sources of spin splitting proportional to the in-plane
hole wavevector were examined briefly. Here, we present
quantitative calculations of spin splitting based on the Luttinger
Hamiltonian approach. We take into account the biaxial strain
inherent in the SiGe heterostructures which gives rise to stronger
splitting of the hole subbands. We consider both the cases of
tensile and compressive strain resulting in the light-hole
(LH)-like and heavy-hole (HH)-like ground state, respectively.
Based on the results for spin splitting, we analyze the
spin-relaxation process for 2D holes. For asymmetric QWs, it is
found that the D'yakonov-Perel' mechanism is more important than
the Elliott-Yafet mechanism. According to our calculations, the
hole spin mean-free path in a strained Si QW can be as large as a
micron at room temperature. This estimate suggests the feasibility
of p-type spintronic applications based on Si, particularly in
view of recent advances in group-IV magnetic
semiconductors.\cite{ge-dms}

\section{Basic equations}

First, we need clarify what we mean by the term "spin". Due
to the degeneracy at the zone center, the spin-orbit interaction
leads to strong hybridization of the hole states and the total
angular momentum must be considered.  If some asymmetry is
introduced such as the confinement in a QW, the degeneracy between
the LH and HH states is lifted at least partly. When the resulting
doublet is well separated energetically from the other states, it
can be treated by using a spin Hamiltonian with the effective spin
$1/2$. So, in speaking of spin states, we mean the states of this
quasi-degenerate doublet.

Let us describe the model used in our calculations. We consider
the case of a QW grown along the $[001]$ direction that is subject
to biaxial strain.  We start from the following $6\times 6$
effective-mass Hamiltonian:
\begin{equation}
\label{eq:1}
H=H_L^{(0)} + H_L^{(\|)} + H_\epsilon  + U(z) I_6.
\end{equation}
Here $H_L \equiv H_L^{(0)} + H_L^{(\|)}$ is the Luttinger
Hamiltonian with $H_L^{(0)}$ corresponding to the part with
$k_{x,y} =0$, $H_\epsilon$ is the contribution due to the biaxial
strain, $U(z)$ is the confining potential which forms the QW, and
$I_6$ is the $6\times 6$ unity matrix. The explicit expression for
$H$ is not provided since it is readily available in the
literature (see, for example, Ref.~\onlinecite{bir-pikus}). Note,
however, that in the following the top of the HH band is used as
the reference energy. As in $\bf k \cdot p$ theory,
$H_L^{(\|)}$ is treated as a perturbation. Following the
conventional notation,\cite{bir-pikus} we choose the {\em
zeroth-order} (i.e., unperturbed) wavefunctions as
\begin{equation}
\label{eq:2}
\Psi_{1ln}=\left(
\begin{array}{c}
0\\ \chi_n^{(l1)}\\ 0\\ 0\\ i \chi_n^{(l2)}\\ 0
\end{array}\right),
\Psi_{2ln}=\left(
\begin{array}{c}
0\\ 0\\  \chi_n^{(l1)}\\ 0\\ 0\\ i \chi_n^{(l2)}
\end{array}\right),
\end{equation}
$$
\Psi_{1hn}=\left(
\begin{array}{c}
\chi_n^{(h)}\\ 0\\ 0\\ 0\\ 0\\ 0
\end{array}\right),
\Psi_{2hn}=\left(
\begin{array}{c}
0\\ 0\\ 0\\ \chi_n^{(h)}\\ 0\\ 0
\end{array}\right),
$$
$$
\Psi_{1sn}=\left(
\begin{array}{c}
0\\ \chi_n^{(s1)}\\ 0\\ 0\\ i \chi_n^{(s2)}\\ 0
\end{array}\right),
\Psi_{2sn}=\left(
\begin{array}{c}
0\\ 0\\  \chi_n^{(s1)}\\ 0\\ 0\\ i \chi_n^{(s2)}
\end{array}\right),
$$
where the plane-wave factors $\exp [i(k_x x+ k_y y)]$ are omitted
for simplicity. In Eq.~(\ref{eq:2}), the first subscript of $\Psi$
(i.e., $1$ or $2$) denotes the doubly degenerate states for
$k_{x,y} =0$, $n$ is the subband number, and $l$, $h$, or $s$
represents the LH, HH, or spin-split (SS) states, respectively. In
fact, if the strain is strong enough such that the corresponding
strain energy $E_\epsilon$ is comparable to the spin-orbital gap
$\Delta$ in the bulk material, the so-called LH and SS states are
hybridized even for  $k_{x,y} =0$. Hereafter, we term the states
"LH" and "SS" which take the genuinely light and spin-split nature
in the limit $\Delta \rightarrow \infty$. This hybridization is
evident from the eigenvalue equations for the LH and SS  envelope
functions $\chi$:
\begin{eqnarray}
\label{eq:3}
H_2 \left( \begin{array}{c}
\chi_n^{(x1)} \\
\chi_n^{(x2)}
\end{array}\right) = E \left( \begin{array}{c}
\chi_n^{(x1)} \\
\chi_n^{(x2)}
\end{array}\right) ,
\end{eqnarray}
\begin{eqnarray}
H_2 = \left(
\begin{array}{l l}
\frac{\hbar^2}{2 m_0} (A+B) \frac{d^2 }{dz^2} +E_\epsilon &
- B \frac{\hbar^2}{\sqrt{2} m_0}\frac{d^2 }{dz^2}  - \frac{E_\epsilon}{\sqrt{2}}  \\
- B \frac{\hbar^2}{\sqrt{2} m_0}\frac{d^2 }{dz^2}  - \frac{E_\epsilon}{\sqrt{2}}  &
\frac{\hbar^2}{2 m_0} A \frac{d^2 }{dz^2}  -\Delta  + \frac{E_\epsilon}{2}
\end{array}\right)  +U(z) I_2.\nonumber
\end{eqnarray}
Here $x$ stands for $l$ or $s$, $A$ and $B$ are parameters of the
hole spectrum,\cite{schaeffler} $m_0$ is the free electron mass,
$I_2$ is the $2\times 2$ unitary matrix, and the deformation
energy mentioned previously is $E_\epsilon = b (2 u_{zz} - u_{xx}
-u_{yy})$, where $b$ is the deformation potential constant and $u$
is the strain tensor. The envelope functions for HH states
$\chi_n^{(h)}$  are determined by the conventional Schr\"{o}dinger
equation with an effective mass $m^{(h)} = m_0/(A-B)$.

In general, further steps require numerical solution of
Eq.~(\ref{eq:3}) to obtain the spectrum and envelope functions of
LH and SS states. Considerable simplification is possible,
however, if the quantization energy is much less than $E_\epsilon$
or $\Delta$, which is often the case for strained SiGe QWs. Under
this approach, it is adequate to use an "effective mass"
approximation for the solution of Eq.~(\ref{eq:3}), which treats
the nondiagonal part of $H_2$ containing the $z$-derivatives as a
perturbation. Subsequently, we obtain
\begin{eqnarray}
\label{eq:5}
\chi_n^{(l1)} = t_{11} \chi_{nl} +t_{12} \sum_{n'} \chi_{n's}
\frac{w_{n'snl}}{E_n^{(l)} - E_{n'}^{(s)}}, \\
\chi_n^{(l2)} = t_{21} \chi_{nl} +t_{22} \sum_{n'} \chi_{n's}
\frac{w_{n'snl}}{E_n^{(l)} - E_{n'}^{(s)}}, \nonumber \\
\chi_n^{(s1)} = t_{12} \chi_{ns} +t_{11} \sum_{n'} \chi_{n'l}
\frac{w_{n'lns}}{E_n^{(s)} - E_{n'}^{(l)}}, \nonumber \\
\chi_n^{(s2)} = t_{22} \chi_{ns} +t_{21} \sum_{n'} \chi_{n'l}
\frac{w_{n'lns}}{E_n^{(s)} - E_{n'}^{(l)}} \nonumber.
\end{eqnarray}
Here the spectrum $E_n^{(l,s)}$ and the envelope functions
$\chi_{nl}$, $\chi_{ns}$ are determined by the conventional
Schr\"{o}dinger equation with the effective masses
\begin{eqnarray}
\label{eq:6}
m^{(l)} = m_0 \left( A + B \left[ \frac{1}{2} +
\frac{9/4+\Delta/(2 E_\epsilon)}{\sqrt{9/4+\Delta/E_\epsilon +
(\Delta/E_\epsilon)^2}}\right]\right)^{-1}, \\
m^{(s)} = m_0 \left( A - B \left[
\frac{9/4+\Delta/(2 E_\epsilon)}{\sqrt{9/4+\Delta/E_\epsilon +
(\Delta/E_\epsilon)^2}}- \frac{1}{2} \right]\right)^{-1}, \nonumber
\end{eqnarray}
and
\begin{eqnarray}
\label{eq:7}
w_{n\alpha n' \beta} = - \frac{\hbar^2}{2 m^\ast}
\int dz\, \chi_{n\alpha} \frac{d^2 \chi_{n'\beta}}{d z^2},\\
m^\ast = m_0 \frac{\sqrt{9 E_\epsilon^2/4+ \Delta E_\epsilon +\Delta^2}}{\sqrt{2}
\Delta B}. \nonumber
\end{eqnarray}
In Eq.~(\ref{eq:5}), $t_{ij}$ are the elements of the unitary
transformation matrix $T$ that approximately diagonalizes $H_2$
(for $U=0$ and $d/dz =0$):
\begin{equation}
\label{eq:8} T= \left( \begin{array}{c c}
\frac{1}{\sqrt{N_l}} & \frac{1}{\sqrt{N_s}} \\
\sqrt{\frac{2}{\sqrt{N_l}}} \left(1 - E^{(l)}/E_\epsilon\right) &
\sqrt{\frac{2}{\sqrt{N_s}}} \left(1 - E^{(s)}/E_\epsilon\right)
\end{array} \right),
\end{equation}
where $E^{(l,s)}$ are the LH and SS states in a strained material
with no confinement (i.e., $U=0$):
\begin{equation}
\label{eq:9}
E^{(l,s)} = \frac{1}{2} \left( \frac{3}{2} E_\epsilon - \Delta \pm
\sqrt{9 E_\epsilon^2/4 + \Delta E_\epsilon +\Delta^2}\right),
\end{equation}
and
\begin{equation}
\label{eq:10}
N_{l,s} = 1+2\left(1-E^{(l,s)}/E_\epsilon\right)^2.
\end{equation}

These basis functions result in nondiagonal terms in the
Hamiltonian $H$. Using the perturbation method,\cite{bir-pikus} it
is possible to transform it to the quasi-diagonal form in any
desired order in $H_L^{(\|)}$. Under this approach, the effective
Hamiltonian for the quasi-degenerate ground subband can be written
as
\begin{equation}
\label{eq:4} H_{eff}^{(l,h)} = -\frac{\hbar^2 (k_x^2+k_y^2)}{2
m_\|^{(l,h)}} I_2 - \frac{\hbar}{2} {\bm \sigma \bm \Omega}^{(l,h)}
(k_x, k_y),
\end{equation}
where $\sigma$ are Pauli matrices. Restricting $\bm \Omega$ to the
terms proportional to $k$ and $k^3$ and neglecting the corrections
to the longitudinal effective mass, we obtain for the ground
LH-like state
\begin{eqnarray}
\label{eq:11} m_\|^{(l)} = m_0 \left[ A - B (\lambda
_{00}^{l1l1}/2 -
\sqrt{2} \lambda_{00}^{l1l2})\right]^{-1}, \\
{\bm \Omega}^{(l)} = {\bm \Omega}^{(l)}_1 + {\bm \Omega}^{(l)}_3, \nonumber \\
{\bm \Omega}^{(l)}_{1x,y} = \frac{\hbar}{m_0} \sqrt{6(3 B^2 +C^2)}
\kappa_{00}^{(l1l2)} k_{x,y}, \nonumber \\
{\bm \Omega}^{(l)}_{3x} = \Pi B k_x (k_x^2+k_y^2) +
\Theta [3 Bk_x (k_x^2 - k_y^2) +2 \sqrt{3 (3 B^2 +C^2)} k_y^2 k_x], \nonumber\\
{\bm \Omega}^{(l)}_{3y} = \Pi B k_y (k_x^2+k_y^2) + \Theta [-3 B
k_y (k_x^2 - k_y^2) +2 \sqrt{3 (3 B^2 +C^2)} k_x^2 k_y] \nonumber,
\end{eqnarray}
where $C$ is a parameter of the hole spectrum\cite{schaeffler} and
\begin{eqnarray}
\label{eq:12}
 \Pi = \frac{\hbar^3}{m_0^2}
\sqrt{\frac{3(3B^2+C^2)}{2}} \sum_{n=0}^{\infty}
\frac{1}{E_0^{(l)} -E_n^{(s)}} \left( \kappa_{0n}^{(l1s2)}
-\kappa_{0n}^{(l2s1)}\right)
\left[\sqrt{2}\left(\lambda_{0n}^{(l1s2)}
+\lambda_{0n}^{(l2s1)}\right) -\lambda_{0n}^{(l1s1)}\right],
\nonumber \\
 \Theta = \frac{\hbar^3}{m_0^2}
\sqrt{\frac{3B^2+C^2}{3}} \sum_{n=0}^{\infty} \frac{1}{E_0^{(l)}
-E_n^{(h)}} \left( \kappa_{0n}^{(l1h)} -\frac{1}{\sqrt{2}}
\kappa_{0n}^{(l2h)}\right) \left(\sqrt{2}\lambda_{0n}^{(l2h)}
+\lambda_{0n}^{(l1h)}\right).
\end{eqnarray}
In Eqs.~(\ref{eq:11}) and (\ref{eq:12}), we use the following
overlap integrals:
\begin{equation}
\label{eq:13} \lambda_{nn'}^{(\alpha\beta)} = \int dz
\chi_n^{(\alpha)} \chi_{n'}^{(\beta)}, \,\,\,\,\,
\kappa_{nn'}^{(\alpha\beta)} = \int dz \chi_n^{(\alpha)} \frac{d
\chi_{n'}^{(\beta)}}{dz}.
\end{equation}
For the ground HH state, we have
\begin{eqnarray}
\label{eq:14}
m_\|^{(h)} = m_0 \left( A + B/2 \right)^{-1}, \\
{\bm \Omega}^{(h)} =  {\bm \Omega}^{(h)}_3, \nonumber \\
{\bm \Omega}^{(h)}_{3x} = \Lambda\left[3 B k_x (k_x^2-k_y^2) +
2 \sqrt{3 (3 B^2 +C^2)} k_y^2 k_x\right], \nonumber\\
{\bm \Omega}^{(h)}_{3y} = \Lambda \left[3 B k_y (k_x^2-k_y^2) + 2
\sqrt{3 (3 B^2 +C^2)} k_x^2 k_y\right] \nonumber,
\end{eqnarray}
where
\begin{eqnarray}
\label{eq:15}
\Lambda = \frac{\hbar^3}{m_0^2} \sqrt{\frac{3B^2+C^2}{3}}
\left[ \sum_{n=0}^{\infty} \frac{1}{E_0^{(h)} -E_n^{(l)}}
\left( \frac{1}{\sqrt{2}} \kappa_{0n}^{(hl2)} -\kappa_{0n}^{(hl1)}\right)
\left(\sqrt{2}\lambda_{0n}^{(hl2)} +\lambda_{0n}^{(hl1)}\right)+ \right.\nonumber \\
\left. \sum_{n=0}^{\infty} \frac{1}{E_0^{(h)} -E_n^{(s)}}
\left( \frac{1}{\sqrt{2}} \kappa_{0n}^{(hs2)} - \kappa_{0n}^{(hs1)}\right)
\left(\sqrt{2}\lambda_{0n}^{(hs2)} +\lambda_{0n}^{(hs1)}\right)\right].
\end{eqnarray}
For both the LH-like and HH-like states, the $z$ component of $\bm
\Omega$ is zero. It is important to note that $\bm \Omega$ is
proportional to the product $\int dz\, \chi d^2 \chi'/dz^2 \, \int
dz\, \chi d\chi'/dz$ for the terms proportional to $k$ and to the
product $\int dz\, \chi \chi' \, \int dz\, \chi d\chi'/dz$ for
those proportional to $k^3$, where $\chi$ and $\chi'$ are the
envelope functions obtained as a solution of Schr\"{o}dinger
equation with the potential $U(z)$ and appropriate effective
masses. For a symmetric $U(z)$, the products of this kind are zero
and the spin splitting vanishes in accordance with the general
symmetry requirements.

Based on these results for hole spectrum, we can calculate the
related spin relaxation rate. Basically, two mechanisms of
relaxation must be addressed. The first is the
D'yakonov-Perel' mechanism.\cite{dp} It is related to spin
precession with a frequency ${\bm \Omega}(k_x,k_y)$, which changes
randomly due to rapid electron transitions in the momentum space.
For this mechanism, the spin relaxation times for the $x$, $y$,
and $z$ spin components obey the following relation  $T_x = T_y =
2 T_z\equiv T_{DP}$.  For non-degenerate carriers, it is given as
\begin{equation}
\label{eq:16}
\frac{1}{T_{DP}} =
\frac{\sum_{k_x,k_y} \Omega_x^2 (k_x,k_y) f_0 (k_x,k_y)}{\sum_{k_x,k_y} f_0 (k_x,k_y)} \tau.
\end{equation}
In this equation, $f_0$ is the Boltzmann distribution function and
$\tau$ is an average characteristic time of electron
scattering;\cite{dp} in the following, we will assume it to be
equal to the electron momentum relaxation time. Another dominant
source of spin relaxation is the Elliott-Yafet mechanism. It is
caused by hybridization of the hole states at the finite
wavevectors.  Using the results for hole wavefunctions, an
order-of-magnitude estimate in a strongly asymmetric QW is
obtained for the ratio of Elliott-Yafet relaxation time $T_{EY}$ to
$T_{DP}$:
\begin{equation}
\label{eq:17}
\frac{T_{EY}}{T_{DP}} \sim \left(\frac{E_Q \tau}{\hbar}\right)^2,
\end{equation}
where $E_Q$ is the characteristic energy of hole confinement, typically of the order
of few tens of meV. The
expression suggests that the D'yakonov-Perel' mechanism provides
the major contribution to spin relaxation in the case of
well-defined quantized energy levels.  Of course, this is not true
for weakly asymmetric QWs, where the D'yakonov-Perel' mechanism
can become less important.

\section{Results and Disucssion}

As a specific example, we consider SiGe inversion layers. The
first case corresponds to a thin Ge layer grown on SiGe, and the
second a Si layer on SiGe; an insulator is placed on top in each
case. Such structures have been studied extensively due to their
perspective applications in p-type enhanced-mobility
metal-oxide-semiconductor field-effect
transistors.\cite{schaeffler} Hole confinement in these devices is
achieved by a strong perpendicular electric field and the
insulator as schematically illustrated in Fig.~1. The barrier at
the left-hand side is assumed to be infinitely high. The
deformation energy in this case is $E_\epsilon = - 2b \left( 2
c_{12}/c_{11} +1\right)\delta$, where $c_{11}$ and $c_{12}$ are
the elastic constants of the QW material (i.e., Si or Ge), $b$ is
the deformation potential constant, and $\delta$ is the relative
mismatch in the lattice parameters. For the case of a relaxed SiGe
buffer, the ground state in the Si QW is LH-like with $E_\epsilon
> 0$, while the Ge QW with $E_\epsilon <0$ has a HH-like ground
state. Note that the band discontinuity at the Si/SiGe or Ge/SiGe
interface is not a well-defined quantity since not only the band
gap but also the spin-orbit energy $\Delta$ are different across
the interface. In our calculations, we assume that the confining
electric field is strong enough to shift the holes away from the
interface.

In Figs.~3 and 4, we show the obtained spin splitting and spin
relaxation rate $1/T_{DP}$ as a function of the confining electric
field at $T=300$ K. Figure~3 presents the case of a
Si/Si$_{0.7}$Ge$_{0.3}$ structure, while Fig.~4 is for
Ge/Si$_{0.3}$Ge$_{0.7}$. The momentum relaxation time $\tau$ of
$5\times 10^{-14}$ s is used, which corresponds roughly to the
reported values of hole mobility.\cite{schaeffler,simob} The
material parameters are taken from Ref.~\onlinecite{schaeffler}.
The spin splitting is calculated for the absolute value of the
wavevector corresponding to the in-plane kinetic energy of $k_B T$
along the $[100]$ (dotted line) or $[110]$ (dashed line)
direction, respectively. For the case of a Si QW, the main
contribution to $\Omega$ at room temperature is due to the terms
proportional to $k^3$.  This is because the $k$-linear term is
proportional to $B$, which is small in the case of Si. On the
other hand, the main contribution to $\Omega$ in the Ge QW is from
the HH$-$LH coupling since the spin-orbital gap $\Delta$ in Ge is
large. Note that the mixing between valence and conduction
band states can noticably influence spin splitting in the latter case (i.e.,
the Ge QW).\cite{holephotogal} For a Si QW, this is less likely to
be important since the direct band gap is 3.5~eV, much larger than
that for Ge (0.9~eV).\cite{madelung}

As can be seen for the figures, the valence band of a Ge QW is
characterized by spin splitting comparable to that for electrons
in III-V QWs. Simultaneously, the hole spin relaxation time is
quite short. In contrast, spin splitting and the relaxation rate
for holes in a strained Si QW are much smaller. The reason for
this is a weak spin-orbit interaction in Si, which is
characteristic for light elements. In particular, the $k^3$
contribution to $\Omega$ is proportional to $\Delta$, which is  as
little as 44~meV for Si. Hence, it can be suggested that by
applying a moderate longitudinal electric field, hole spin in a
strained Si QW can be transferred over a distance of a micron or
so without the loss of coherence. On the other hand, spin
relaxation can be tuned effectively by modulating the confining
electric field. It is important to note once more that in unstrained
structures, say conventional $p$-type Si/SiO$_2$ inversion channels,
spin relaxation is
expected to be much faster due to relatively small subband
separation and, as a result, strong hybridization of the hole states
at finite wavevectors.

\section{Conclusion}
We predict that spin splitting of 2D holes and the related spin
relaxation time in strained SiGe QWs can be effectively controlled
by the structure composition and the degree of QW asymmetry. For
field-effect transistor-like structures,\cite{tanaka} spin
splitting depends strongly on the value of the perpendicular
electric field. In particular, the hole spin relaxation time in a
strained Si QW channel can be in the hundred picosecond range at
room temperature, making it suitable for spintronic applications.

\begin{acknowledgments}
This work was supported in part by the Defense Advanced Research
Projects Agency, the SRC/MARCO Center on FENA, and the CRDF grant
UE2-2439-KV-02.
\end{acknowledgments}

\clearpage

\begin{figure}
\includegraphics[scale=0.5]{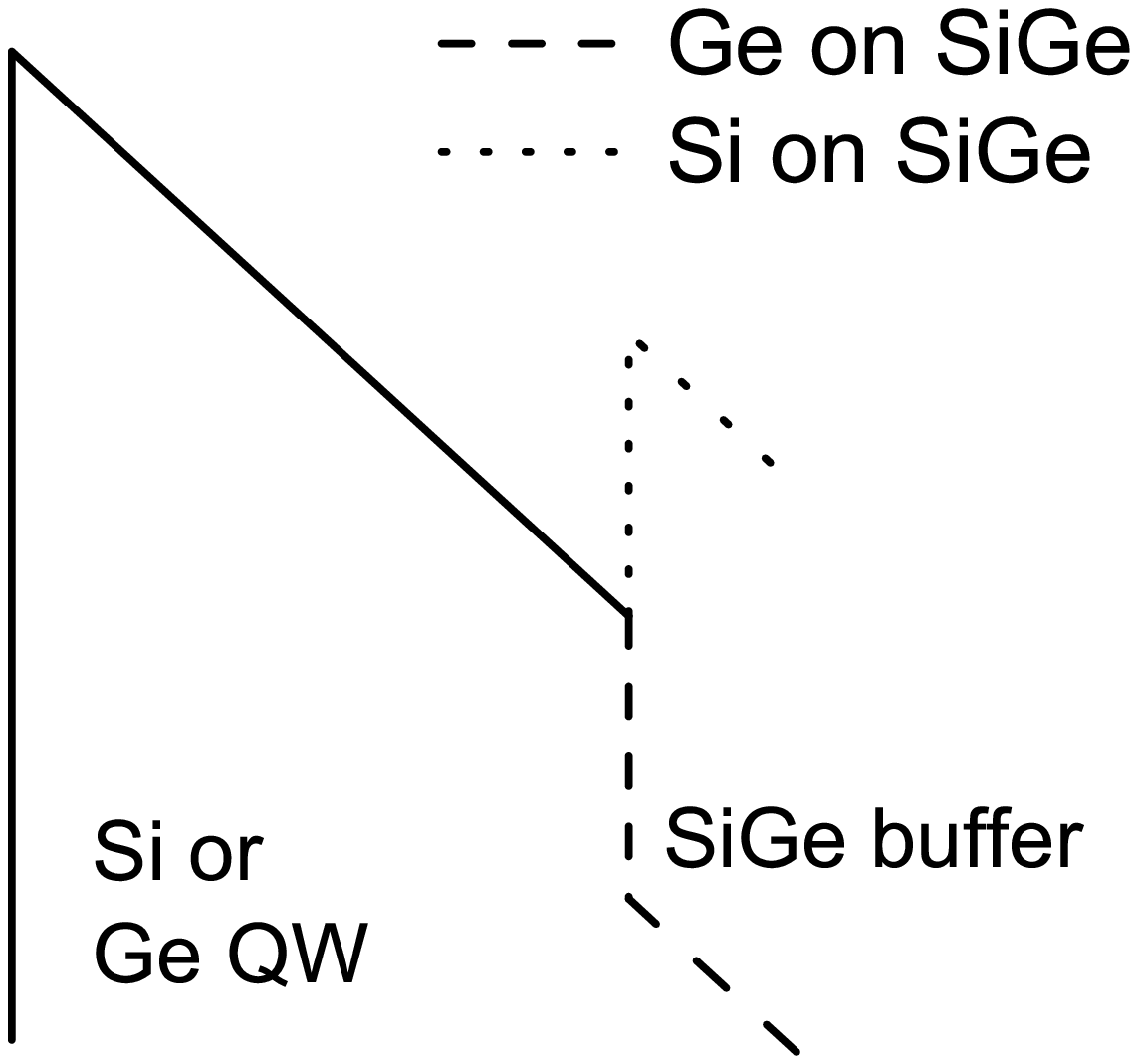}
\caption{ Model valence band profile used for spin splitting
calculations. Holes are assumed to be confined in a triangular
well with an infinite barrier on the left-hand side. The band
discontinuity at the SiGe interface depends on the structure type
(Si/SiGe or Ge/SiGe) and was disregarded. }
\end{figure}

\clearpage
\begin{figure}
\includegraphics[scale=0.5]{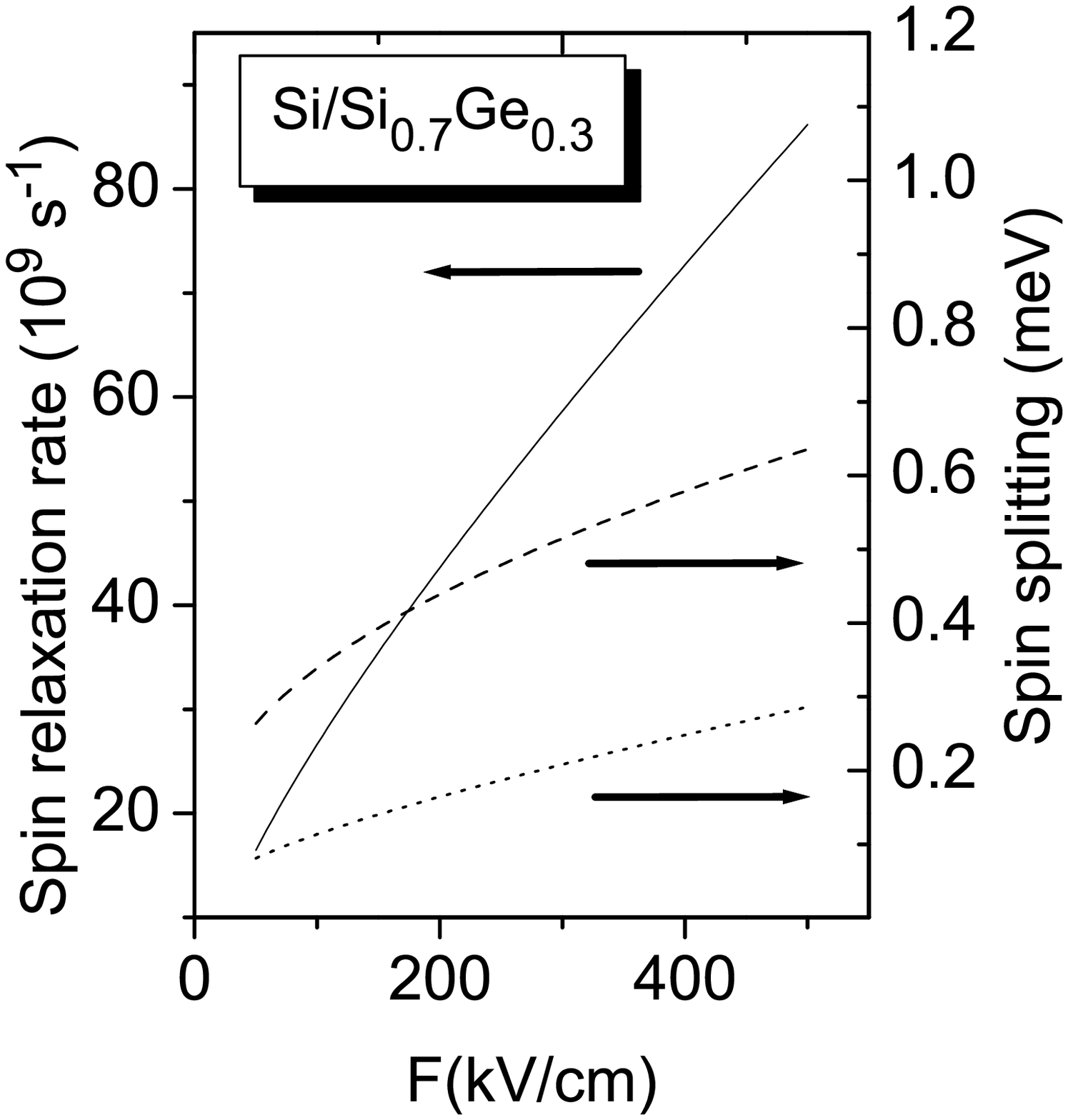}
\caption{ Spin splitting and D'yakonov-Perel' spin relaxation rate
for a strained Si QW grown on the Si$_{0.7}$Ge$_{0.3}$ buffer.
Spin splitting is given for the hole kinetic energy of $k_B T$
along the $[100]$ (dotted line) and $[110]$ (dashed line)
directions. }
\end{figure}

\clearpage
\begin{figure}
\includegraphics[scale=0.5]{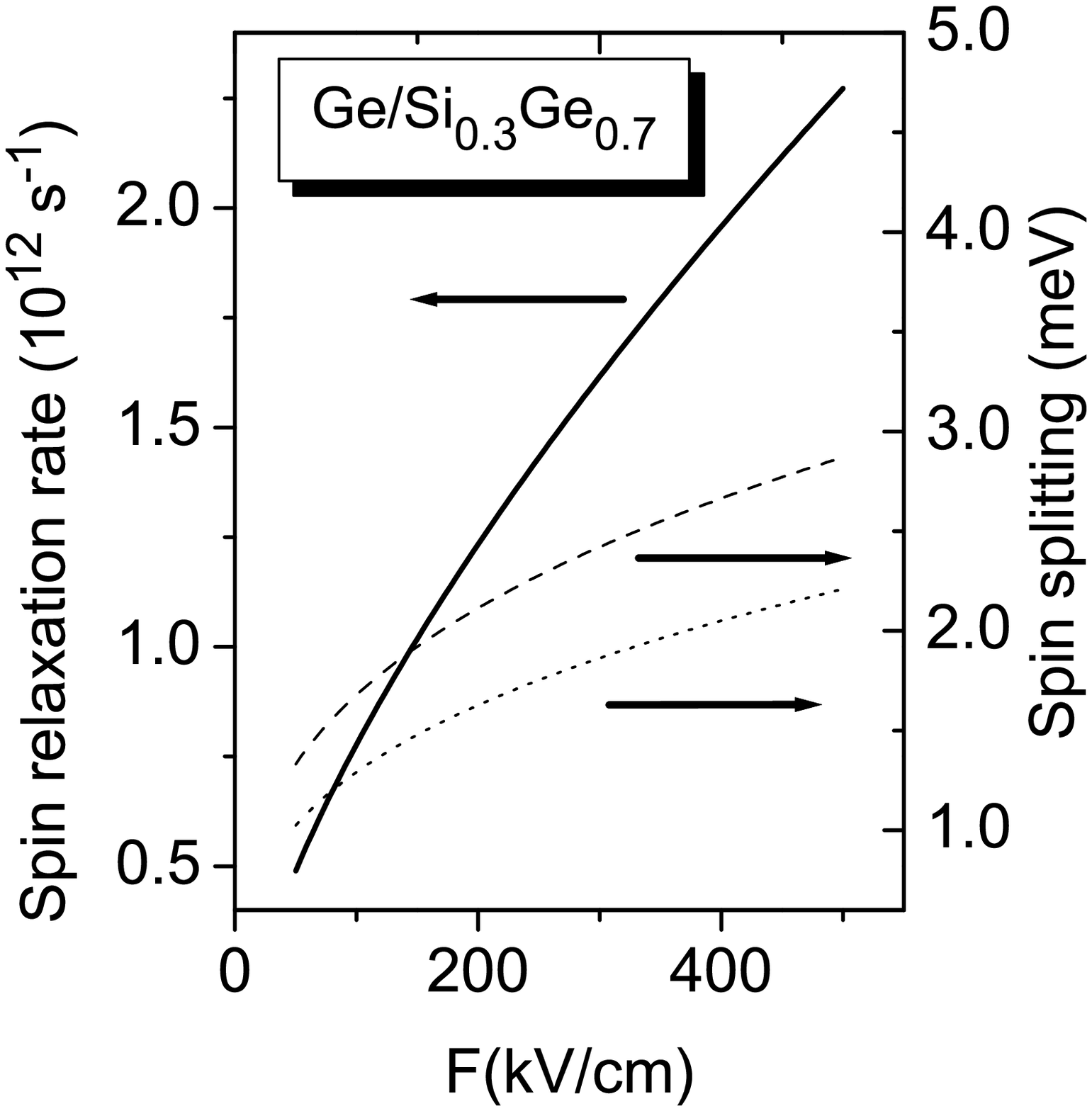}
\caption{ Spin splitting and  D'yakonov-Perel' spin relaxation
rate for a strained Ge QW grown on the Si$_{0.3}$Ge$_{0.7}$
buffer. Spin splitting is given for the hole kinetic energy of
$k_B T$ along the $[100]$ (dotted line) and $[110]$ (dashed line)
directions.}
\end{figure}

\end{document}